\documentclass[draftclsnofoot,onecolumn]{IEEEtran}
\usepackage{fullpage}
\usepackage{citesort}

\usepackage{epsfig}
\usepackage{graphicx}
\usepackage{caption}
\usepackage{subcaption}
\usepackage{fancybox}
\usepackage{color}

\usepackage[cmex10]{amsmath}
\usepackage{amsfonts}
\usepackage{cases}
\usepackage{url}

\usepackage{epstopdf}
\usepackage{amsthm}

\graphicspath{{./} {img/}}
\newtheorem{THEO}{Theorem}
\newtheorem{CLAI}{Claim}

\newtheorem{CONJ}{Conjecture}
\newtheorem*{REMA}{Remark}

\begin{document}
%
\title{
\vspace*{-5mm}
Mixture Gaussian Signal Estimation \\with $\ell_\infty$ Norm Error}
%
%
%

\author{Jin~Tan,~\IEEEmembership{Student Member,~IEEE,}
Dror~Baron,~\IEEEmembership{Senior Member,~IEEE,}
~and~Liyi~Dai
\thanks{This work was supported by the grants NSF CCF-1217749 and ARO W911NF-04-D-0003. Portions appeared at the Information Theory and Application workshop (ITA), Feb. 2013~\cite{TB2013ITA}.}
\thanks{Jin Tan and Dror Baron are with the Department of Electrical and Computer Engineering, North Carolina State University, Raleigh, NC 27695. E-mail: \{jtan,barondror\}@ncsu.edu. Liyi Dai is with the Computing Sciences Division, US Army Research Office. E-mail:liyi.dai.civ@mail.mil.}
}
\maketitle \thispagestyle{empty}
\newcommand{\xhat}{\widehat{\mathbf{x}}}
\newcommand{\xhati}{\widehat{x}_i}
\def\x{{\mathbf x}}
\def\L{{\cal L}}
\begin{abstract}
We consider the problem of estimating an input signal from noisy measurements in both parallel scalar Gaussian channels and linear mixing systems. The performance of the estimation process is quantified by the~$\ell_\infty$ norm error metric. We first study the minimum mean~$\ell_\infty$ error estimator in parallel scalar Gaussian channels, and verify that, when the input is independent and identically distributed (i.i.d.) mixture Gaussian, the Wiener filter is asymptotically optimal with probability~$1$. 
For linear mixing systems with i.i.d. sparse Gaussian or mixture Gaussian inputs, 
under the assumption that the relaxed belief propagation (BP) algorithm matches Tanaka's fixed point equation,
applying the Wiener filter to the output of relaxed BP is also asymptotically optimal with probability~$1$.
However, in order to solve the practical problem where the signal dimension is finite, we apply an estimation algorithm that has been proposed in our previous work, and illustrate that an $\ell_\infty$ error minimizer can be approximated by an $\ell_p$ error minimizer provided the value of $p$ is properly chosen.
\end{abstract}

\begin{IEEEkeywords}
Belief propagation, estimation theory, $\ell_\infty$ norm error, linear mixing systems, parallel scalar channels.
\end{IEEEkeywords}

%
\IEEEpeerreviewmaketitle

\section{Introduction}
\label{sec:intro}
\subsection{Motivation}
\label{subsec:motivation}
The Gaussian distribution is widely used to describe the probability densities of various types of data, owing to its mathematical advantages~\cite{Papoulis91}. It has been shown that non-Gaussian distributions can often be represented by an infinite mixture of Gaussian~\cite{Alecu2006}, so that the mathematical advantages of the Gaussian distribution can be preserved when discussing non-Gaussian data models~\cite{Alecu2005,Bijaoui2002}. 

A set of parallel scalar Gaussian channels with a mixture Gaussian input vector has been used to model image denoising problems
~\cite{Tabuchi2007,Alecu2005,Bijaoui2002},
while linear mixing systems are popular models used in many settings such as compressed 
sensing~\cite{CandesRUP,DonohoCS}, regression~\cite{Huber1973,OLeary1990}, and multiuser detection~\cite{GuoWang2008}.
Signal reconstruction from noisy measurements is prevalent in the literature, but the
minimization of the $\ell_{\infty}$ error has received less attention.
Our interest in the $\ell_{\infty}$ error is motivated by applications such as
group testing~\cite{Gilbert2012} and trajectory planning in control systems~\cite{Egerstedt1999},
where we want to decrease the worst-case sensitivity to noise.

\subsection{Problem setting}
\label{subsec:Prob}

We describe {\em parallel scalar Gaussian channels} and {\em linear mixing systems} below. In both settings, the input vectors~$\x$ are independent and identically distributed (i.i.d.)
 mixture Gaussian, i.e.,~$x_i\sim s_1\cdot\mathcal{N}(0,\mu_1)+s_2\cdot\mathcal{N}(0,\mu_2)+\cdots+s_K\cdot\mathcal{N}(0,\mu_K)$, where~$s_1,s_2,\ldots,s_K\ge0$ are given,~$\sum_{k=1}^Ks_k=1$, and~$\mu_1,
\mu_2,\ldots,\mu_K$ are also given. The subscript~$(\cdot)_i$ denotes the $i$-th element of the 
corresponding vector. As a specific case, we study the i.i.d.~$s$-sparse Gaussian, i.e.,~$x_i\sim s\cdot\mathcal{N}(0,\mu_x)+(1-s)\delta_0(x_i)$ for some given~$s$ and~$\mu_x$.

For a set of {\em parallel scalar Gaussian channels}~\cite{Tabuchi2007,Alecu2005,Bijaoui2002}, we consider
\begin{equation}
{\bf r=x+z},
\label{eq:scalar}
\end{equation}
where~${\bf r,x,z}\in\mathbb{R}^N$. The vectors~${\bf r}$,~${\bf x}$, and~${\bf z}$ are the received signal, the input signal, and the i.i.d. Gaussian noise, respectively. The additive Gaussian noise channel can also be described by the conditional distribution
\begin{equation}
f_{\bf R|X}({\bf r|x})=\prod_{i=1}^Nf_{R|X}(r_i|x_i)=\prod_{i=1}^N\frac{1}{\sqrt{2\pi\mu_z}}\exp{\left(-\frac{(r_i-x_i)^2}{2\mu_z}\right)},
\label{eq:Gchannel}
\end{equation}
where~$\mu_z$ is the variance of the Gaussian noise.

For a linear system~\cite{CandesRUP,DonohoCS,GuoWang2008},
\begin{equation}
\label{eq:basicSystem}
\mathbf{w=\Phi x},
\end{equation}
the random linear mixing matrix (or measurement matrix)~$\mathbf{\Phi}\in\mathbb{R}^{M\times N}$ is known and its entries are i.i.d. Because each component in the measurement vector~$\mathbf{w}\in\mathbb{R}^M$ is a linear combination of the components in~$\mathbf{x}$, we call the system~\eqref{eq:basicSystem} a {\em linear mixing system}. The measurements~$\mathbf{w}$ are passed through a bank of separable scalar channels characterized by conditional distributions,
\begin{eqnarray}
f_{\mathbf{Y|W}}(\mathbf{y|w})=\prod_{i=1}^M f_{Y|W}(y_i|w_i),
\label{eq:DisChannel}
\end{eqnarray}
where $\mathbf{y}$ is the channel output vector. {However, unlike the parallel scalar Gaussian channels~\eqref{eq:Gchannel}, the channels~\eqref{eq:DisChannel} for the linear mixing system are general and are not restricted to Gaussian \cite{GuoWang2007,Rangan2010}}. 

Our goal is to reconstruct the original system input $\mathbf{x}$ from the channel output~${\bf r}$~\eqref{eq:scalar} or from the output~$\mathbf{y}$ and the matrix $\mathbf{\Phi}$~\eqref{eq:basicSystem}. 
To evaluate how accurate the reconstruction process is, we calculate the error between the original signal $\mathbf{x}$ and the reconstructed signal $\xhat$. Many works emphasize the squared error performance~\cite{GuoVerdu2005,CSBP2010,TroppOMP}. In this paper, however, we focus on preventing any significant errors during the signal reconstruction process. That is, we want to study algorithms that minimize the $\ell_\infty$ norm of the error,
\begin{equation*}
\|\widehat{\mathbf{x}}-\mathbf{x}\|_{\infty} = 
\max_{i \in \{1,\ldots,N\} } |\widehat{x}_i - x_i|.
\end{equation*}

\subsection{Related work}
\label{subsec:relatedWork}

In our previous work \cite{tan2012signal}, we dealt with an additive error metric defined as
\begin{eqnarray}
D(\mathbf{\widehat{x},x})=\sum_{i=1}^N d(\widehat{x}_i, x_i).
\label{eq:distDef}
\end{eqnarray}
We proposed a reconstruction algorithm that is optimal in minimizing the expected value of error metrics of the form \eqref{eq:distDef}, where the reconstruction process is done component-wise, i.e., for each $i\in\{1,2,...,N\}$, $d(\widehat{x}_i,x_i)$ is minimized separately. However, in contrast to \eqref{eq:distDef}, the~$\ell_\infty$ error is not additive, because it only considers the one component that has the maximum absolute value, and thus it is not straightforward to extend the algorithm \cite{tan2012signal} to minimize the $\ell_\infty$ error.

There have been a number of studies on general properties of $\ell_\infty$ error related solutions. An overdetermined linear system $\mathbf{y=\Phi x}$, where $\mathbf{\Phi}\in\mathbb{R}^{N\times M}$ and $N>M$, was considered by Cadzow~\cite{cadzow2002}, and the properties of the minimum $\ell_\infty$ error solutions to this system were explored. In Clark~\cite{clark1961}, the author developed a way to calculate the distribution of the greatest element in a finite set of random variables. And in Indyk~\cite{indyk2001}, an algorithm was introduced to find the nearest neighbor of a point while $\ell_\infty$ norm distance was considered. Finally, Lounici~\cite{Lounici2008} studied the $\ell_\infty$ error convergence rate for Lasso and Dantzig estimators.

\subsection{Contributions}
\label{sec:contrib}
Our first result is asymptotic in nature; we prove that, in parallel scalar Gaussian channels where the input signal is i.i.d. sparse Gaussian or i.i.d. mixture Gaussian, the Wiener filter~\cite{Wiener1949} is asymptotically optimal for $\ell_\infty$ norm error. These results are extended to linear mixing systems based on the assumption that the relaxed BP algorithm~\cite{Rangan2010} matches Tanaka's fixed point equation~\cite{GuoVerdu2005,Montanari2006,GuoBaronShamai2009,RFG2010,DMM2009,RanganGAMP2010}. We claim that in linear mixing systems, when the input signal is i.i.d. sparse Gaussian or i.i.d. mixture Gaussian, applying the Wiener filter to the outputs of the relaxed BP algorithm is asymptotically optimal for~$\ell_\infty$ norm error. 

Our second result is practical in nature; in order to deal with signals of finite length~$N$ in practice, we apply the $\ell_p$ error minimization by~\cite{tan2012signal}, and show numerically that, with a finite signal length~$N$, the $\ell_p$ error minimization~\cite{tan2012signal} outperforms the Wiener filtering.

The remainder of the paper is arranged as follows. We review the metric-optimal algorithm along with the relaxed BP algorithm  in Section \ref{sec:review}, and then discuss our main results in Section \ref{sec:main}. Simulation results are given in Section \ref{sec:NumSim}, and Section~\ref{sec:concld} concludes. Proofs of the main results appear in appendices.

\section{Review of the Relaxed Belief Propagation and Metric-Optimal Algorithms}
\label{sec:review}
The set of parallel scalar Gaussian channels~\eqref{eq:scalar} has a simple structure, because each channel~$r_i=x_i+z_i$ is separable from other scalar channels, and thus the analysis on the system model~\eqref{eq:scalar} is convenient. The linear mixing systems, however, is more complicated. Previous works~\cite{Guo2006,CSBP2010,Rangan2010,GuoVerdu2005,Montanari2006,GuoBaronShamai2009,RFG2010,DMM2009,RanganGAMP2010} have shown that a linear mixing system~\eqref{eq:basicSystem} and~\eqref{eq:DisChannel} can be decoupled to parallel scalar Gaussian channels. In this section, we review the decoupling process of the linear mixing systems, as well as the metric-optimal algorithm that is based on the decoupling.

There are different versions of the relaxed belief propagation (BP) algorithm~\cite{Guo2006,CSBP2010,Rangan2010} in the literature, while our proposed algorithm~\cite{tan2012signal} is based on the one by Rangan~\cite{Rangan2010}, specifically the software package ``GAMP"~\cite{Rangan:web:GAMP}. An important result from the relaxed BP algorithm in linear mixing system problems~\cite{Rangan2010} is that, after a sufficient number of iterations, the relaxed BP process calculates a vector $\mathbf{q}=[q_1,q_2,\ldots,q_N]^T\in\mathbb{R}^N$, and then estimating the inputs $\mathbf{x}$ from the outputs $\mathbf{y}$ of a linear mixing system \eqref{eq:basicSystem} and~\eqref{eq:DisChannel} is asymptotically statistically equivalent to estimating each input entry $x_i$ from the corresponding $q_i$, where $q_i$ is regarded as the output of a scalar Gaussian channel:
\begin{eqnarray}
\label{eq:scalarGchannel}
q_i=x_i+v_i
\end{eqnarray}
for $i\in\{1,\ldots,N\}$,
where each channel's additive noise $v_i$ is Gaussian distributed $\mathcal{N}(0,\mu_v)$, and $\mu_v$ satisfies Tanaka's fixed point equation~\cite{GuoVerdu2005,Montanari2006,GuoBaronShamai2009,RFG2010,DMM2009,RanganGAMP2010}.
The value of $\mu_v$ can also be obtained from the relaxed BP process~\cite{Rangan2010}. 
We note in passing that when we discuss~$q_i=x_i+v_i$, these are the parallel scalar channels resulting from the relaxed BP algorithm, and when we discuss~$r_i=x_i+z_i$, these are the true parallel Gaussian channels~\eqref{eq:scalar}.

\begin{figure}[t]
\centering
\includegraphics[width=8.5cm]{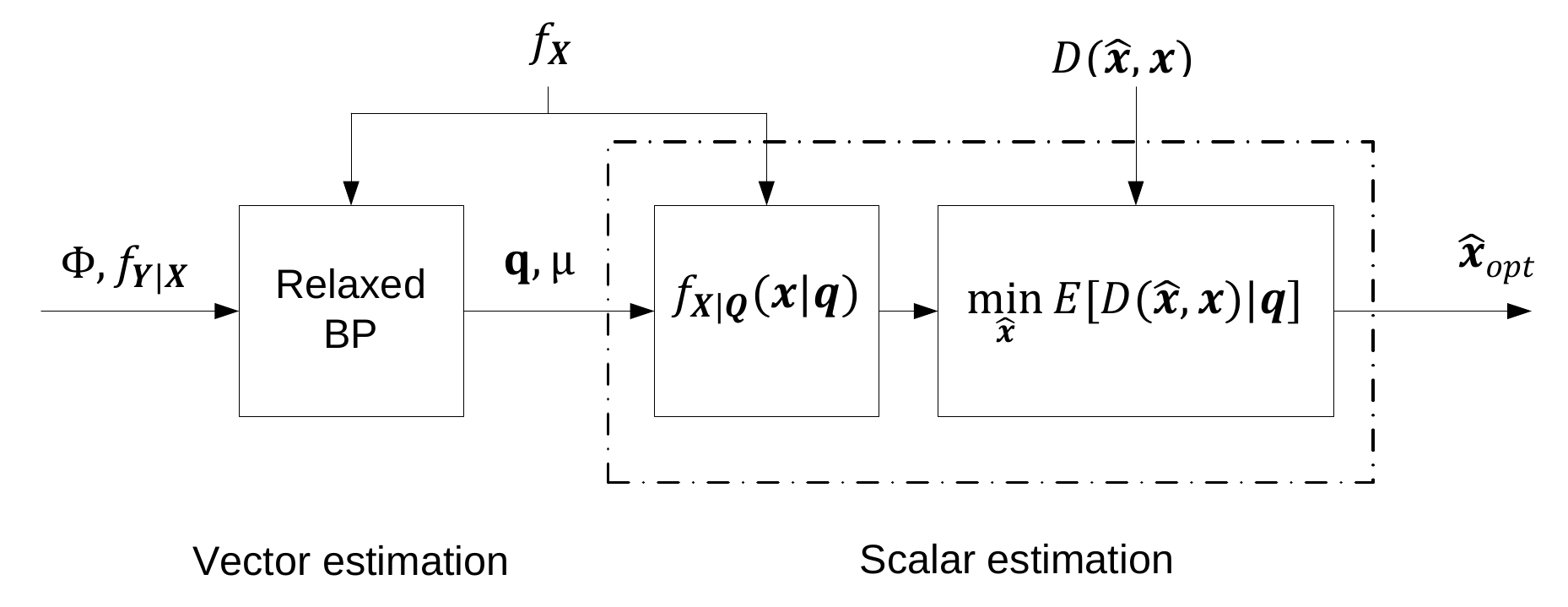}
\vspace*{-5mm}
\caption{
{\small\sl
The structure of the metric-optimal estimation algorithm.
}
\label{fig:EstSys}
}
\vspace*{-5mm}
\end{figure}

In previous related work~\cite{tan2012signal}, we utilized the outputs of the relaxed BP algorithm~\cite{Rangan2010}, and introduced a general metric-optimal estimation algorithm that deals with arbitrary error metrics. Figure~\ref{fig:EstSys} illustrates the structure of our metric-optimal algorithm (dashed box). The algorithm is essentially a scalar estimation process, whereas the relaxed BP algorithm deals with the vector estimation.

We first compute the conditional probability density function $f_{\mathbf{X|Q}}(\mathbf{x|q})$ from Bayes' rule:
\begin{eqnarray}
f_{\mathbf{X|Q}}(\mathbf{x|q})&=& 
\frac{f_{\mathbf{Q|X}}(\mathbf{q|x})f_\mathbf{X}(\mathbf{x})}{\int f_{\mathbf{Q|X}}(\mathbf{q|x})f_\mathbf{X}(\mathbf{x}) d\mathbf{x}}.
\label{eq:BayesRule}
\end{eqnarray}
Then, given an additive error metric $D(\mathbf{\widehat{x},x})$, the optimal estimand $\widehat{\mathbf{x}}_{\text{opt}}$ is generated by minimizing the conditional expectation of the error metric $E[D(\mathbf{\widehat{x},x)|q}]$:
\begin{eqnarray}
\widehat{\mathbf{x}}_{\text{opt}}&=&\arg\min_{\mathbf{\widehat{x}}}E[D(\mathbf{\widehat{x},x)|q}]\nonumber\\
&=&\arg\min_{\mathbf{\widehat{x}}} \int D(\mathbf{\widehat{x},x})f_{\mathbf{X|Q}}(\mathbf{x|q})d\mathbf{x}.
\label{eq:mainAlg}
\end{eqnarray}
In the large system limit, the estimand satisfying \eqref{eq:mainAlg} is asymptotically optimal, because it minimizes the conditional expectation of the error metric. 

Because both the error metric function $D(\mathbf{\widehat{x},x})$ and the conditional probability function $f_{\mathbf{X|Q}}(\mathbf{x|q})$ are separable, the problem reduces to scalar estimation~\cite{Levy2008}. The estimand $\xhat_\text{opt}$ is solved in a component-wise fashion:
\begin{eqnarray}
\widehat{x}_{\text{opt},i}&=&\arg\min_{{\widehat{x}_{i}}}E[D({\widehat{x}_i,x_i)|q_i}]\nonumber\\
&=&\arg\min_{\widehat{x}_i} \int D({\widehat{x}_i,x_i})f({x_i|q_i})d{x_i},
\label{eq:scalarEst}
\end{eqnarray}
for each $x_i$, $i\in\{1,2,3,...,N\}$. This scalar estimation is easy and fast to implement.

Owing to the fact that $\ell_\infty$ error only considers the component with greatest absolute value, and does not have an additive form \eqref{eq:distDef}, it is natural to turn to $\ell_p$ norm error as an alternative.

Recall that the definition of the $\ell_p$ norm error between $\xhat$ and $\mathbf{x}$ is
\begin{equation}
\|\widehat{\mathbf{x}}-\mathbf{x}\|_p=\left(\sum_{i \in \{1,\ldots,N\} } |\widehat{x}_i-x_i|^p\right)^{1/p}.\nonumber
\end{equation}
This type of error is closely related to our definition of the error metric~\eqref{eq:distDef}. We define
\begin{equation}
D_p(\mathbf{\widehat{x},x})=\sum_{i=1}^N |\widehat{x}_i- x_i|^p=\|\mathbf{\widehat{x}-x}\|_p^p\label{eq:metricDef_lp},\nonumber
\end{equation}
and let $\widehat{\mathbf{x}}_p$ denote the estimand that minimizes the conditional expectation of $D_p(\mathbf{\widehat{x},x})$, i.e.,
\begin{eqnarray}
\widehat{\mathbf{x}}_p&=&\arg\min_{\widehat{\mathbf{x}}} E[D_p(\mathbf{\xhat,x})|\mathbf{q}]\nonumber\\
&=&\arg\min_{\widehat{\mathbf{x}}} E[\|\mathbf{\widehat{x}-x}\|_p^p|\mathbf{q}],
\label{eq:xhat_p}
\end{eqnarray}
and
\begin{equation}
\widehat{x}_{p,i}=\arg\min_{\widehat{x}_i} E[|\widehat{x}_i-x_i|^p|q_i],
\label{eq:xhat_pi}
\end{equation}
for $i\in\{1,2,...,N\}$.

Although $\xhat_p$ is minimizing the $(\ell_p)^p$ error, rather than the $\ell_p$ error, we call $\xhat_p$ the {\em minimum mean $\ell_p$ norm error estimator} for simplicity.

Because it can be shown that
\begin{equation}
\lim_{p\rightarrow\infty}\|\widehat{\mathbf{x}}-\mathbf{x}\|_p= \|\widehat{\mathbf{x}}-\mathbf{x}\|_\infty,
\label{eq:p_and_infty}
\end{equation}
it is reasonable to expect that if we set $p$ to a large value, then running our metric-optimal algorithm with error metric $D_p(\cdot)$~\eqref{eq:metricDef_lp} will give a solution that converges to an estimand that minimizes the $\ell_\infty$ error.

\section{Main Results}
\label{sec:main}

In this section, we first study the minimum mean~$\ell_\infty$ error estimator for parallel scalar Gaussian channels~\eqref{eq:scalar}, then discuss the minimum mean~$\ell_\infty$ error estimator for linear mixing systems~\eqref{eq:basicSystem} and~\eqref{eq:DisChannel}, and finally analyze the performance of the minimum mean $\ell_p$ norm error estimators~\eqref{eq:xhat_p} in terms of the $\ell_\infty$ norm error.

\subsection{The minimum mean~$\ell_\infty$ estimator for parallel scalar Gaussian channels}
\label{subsec:scalarG}

For a set of parallel scalar Gaussian channels~\eqref{eq:scalar}, the minimum mean squared error estimator, i.e., $p=2$ in~\eqref{eq:xhat_p} (here we replace~${\bf q}$ by~${\bf r}$ for scalar Gaussian channel discussion), is achieved by the conditional expectation~$E[{\bf x|r}]$. To build intuition into the problem of finding~$E[{\bf x|r}]$, we first suppose for simplicity that~${\bf x}$ is Gaussian (not mixture Gaussian), i.e.,~${\bf x}\sim\mathcal{N}({\bf0},\mu_x\cdot I_N)$ and~${\bf z}\sim\mathcal{N}({\bf0},\mu_z\cdot I_N)$, where~$I_N$ is the~$N\times N$ identity matrix, then the estimand
$
{\bf\widehat{x}}_{2}=E[{\bf x|r}]=\frac{\mu_x}{\mu_x+\mu_z}{\bf r}
$
gives the minimum mean squared error. This format $\frac{\mu_x}{\mu_x+\mu_z}{\bf r}$ is called the {\em Wiener filter} in signal processing~\cite{Wiener1949}.
It has been shown by Sherman~\cite{Sherman1958} that, when the signal input vector and the parallel Gaussian channels are both Gaussian, the linear Wiener filter is also optimal for all  $\ell_p$ norm errors ($p\ge1$), including the~$\ell_\infty$ norm error. Surprisingly, we find that, if the signal input is i.i.d. sparse Gaussian or i.i.d. mixture Gaussian, the Wiener filter asymptotically minimizes the $\ell_\infty$ error. Our main results follow.

\begin{THEO}
\label{thm:01}
In a set of parallel scalar Gaussian channels~\eqref{eq:scalar}, if the input signal~${\bf x}$ is i.i.d. sparse Gaussian, i.e.,~$x_i\sim s\cdot\mathcal{N}(0,\mu_x)+(1-s)\delta_0(x_i)$, then the Wiener filter 
\begin{equation}
{\bf \widehat{x}}_{\text{W,SG}}=\frac{\mu_x}{\mu_x+\mu_z}{\bf r}
\nonumber
\end{equation}
is asymptotically optimal for~$\ell_\infty$ error with probability~$1$. More specifically,
\begin{equation}
\Pr\left\{E\left[\left.\lim_{N\rightarrow\infty}\|{\bf x-\widehat{x}}_\text{W,SG}\|_\infty\right|{\bf r}\right]\le 
E\left[\left.\lim_{N\rightarrow\infty}\|{\bf x-\widehat{x}}\|_\infty\right|{\bf r}\right]\right\}=1,
\nonumber
\end{equation}
where~${\bf \widehat{x}}$ is any arbitrary estimand.
\end{THEO}

Theorem~\ref{thm:01} is proved in Appendix~\ref{appen:ThmProof}. The main idea of the proof is to show that asymptotically the maximum absolute error lies in index~$i$,~$\|{\bf x-\widehat{x}}\|_\infty=|x_i-\widehat{x}_i|$, {\em where $i$ is such that $x_i$ is nonzero}. Therefore, minimizing the maximum absolute error between the estimand~${\bf \widehat{x}}$ and the entire vector~${\bf x}=[x_1,x_2,\ldots,x_N]^T$ is equivalent to minimizing the maximum absolute error between the estimand and the subvector~${\bf\widetilde{x}}=[x_{l_1},x_{l_2},\ldots,x_{l_d}]^T$, where~$x_{l_i}$'s are nonzero and Gaussian distributed and~$d$ represents the number of nonzero elements in~${\bf x}$. 
Thus for an i.i.d. Gaussian vector~${\bf\widetilde{x}}$, the Wiener filter minimizes the~$\ell_\infty$ error~\cite{Sherman1958}.

Theorem~\ref{thm:01} only applies to~$s$-sparse Gaussian signals, but can be easily extended to the mixture Gaussian distribution, thus significantly enhancing the applicability of our result. In the mixture Gaussian input case, the maximum absolute error between~${\bf x}$ and the estimand~${\bf \widehat{x}}$ lies in the index~$i$ that corresponds to the Gaussian mixture component with greatest variance.
\begin{THEO}
\label{coro:01}
In a set of parallel scalar Gaussian channels~\eqref{eq:scalar}, if the input signal~${\bf x}$ is i.i.d. mixture Gaussian, i.e.,~$x_i\sim s_1\cdot\mathcal{N}(0,\mu_1)+s_2\cdot\mathcal{N}(0,\mu_2)+\cdots+ s_K\mathcal{N}(0,\mu_K)$, where~$s_1,s_2,\ldots,s_K\ge0$ and~$\sum_{k=1}^K s_k=1$, then the Wiener filter 
\begin{equation}
{\bf \widehat{x}}_{\text{W,MG}}=\frac{\mu_\text{max}}{\mu_\text{max}+\mu_z}{\bf r}
\nonumber
\end{equation}
is asymptotically optimal for~$\ell_\infty$ error with probability~$1$, where~$\mu_\text{max}=\max_{k\in\{1,2,\ldots,K\}}\mu_k$. More specifically,
\begin{equation}
\Pr\left\{E\left[\left.\lim_{N\rightarrow\infty}\|{\bf x-\widehat{x}}_\text{W,MG}\|_\infty\right|{\bf r}\right]\le 
E\left[\left.\lim_{N\rightarrow\infty}\|{\bf x-\widehat{x}}\|_\infty\right|{\bf r}\right]\right\}=1,
\nonumber
\end{equation}
where~${\bf \widehat{x}}$ is any arbitrary estimand.
\end{THEO}

The proof of Theorem~\ref{coro:01} is given in Appendix~\ref{appen:coro_01}.

\subsection{The minimum mean~$\ell_\infty$ error estimator for linear mixing systems}
\label{eq:infty_lin}
We discussed in Section~\ref{sec:review} that, using the statistical information of the linear mixing system~\eqref{eq:basicSystem}~and~\eqref{eq:DisChannel}, the relaxed BP algorithm asymptotically computes a set of equivalent parallel scalar Gaussian channels. Therefore, using the output of the relaxed BP algorithm, i.e. the scalar Gaussian channels output vector~${\bf q}$ and the noise variance~$\mu_v$, and then applying the Wiener filter, we will obtain the estimand that is asymptotically optimal in the~$\ell_\infty$ error sense for the linear mixing system~\eqref{eq:basicSystem} and~\eqref{eq:DisChannel}. Because the analysis of the equivalent scalar Gaussian channels~\eqref{eq:scalarGchannel} relies on the replica method~\cite{GuoVerdu2005}, which has only been rigorously justified in specific setting~\cite{Bayati2011}, we state our result below as claim.

\begin{CLAI}
Given the system model described by~\eqref{eq:basicSystem} and~\eqref{eq:DisChannel}, where the input signal~${\bf x}$ is i.i.d. mixture Gaussian (sparse Gaussian is a specific case) distributed,~$x_i\sim s_1\cdot\mathcal{N}(0,\mu_1)+s_2\cdot\mathcal{N}(0,\mu_2)+\cdots+ s_K\mathcal{N}(0,\mu_K)$, where~$s_1,s_2,\ldots,s_K\ge0$ and~$\sum_{k=1}^K s_k=1$, as the signal dimension~$N\rightarrow\infty$ and the measurement ratio~$M/N$ is fixed, the estimand
\begin{equation}
\widehat{\mathbf{x}}_{\text{W,MG-BP}}=\frac{\mu_\text{max}}{\mu_\text{max}+\mu_v}{\bf q}\nonumber
\end{equation}
is asymptotically optimal for~$\ell_\infty$ error with probability~$1$, where~${\bf q}$ and~$\mu_v$~\eqref{eq:scalarGchannel} are the outputs of the relaxed BP algorithm, and~$\mu_\text{max}=\max_{k\in\{1,2,\ldots,K\}}\mu_k$.
\end{CLAI}

The relaxed BP algorithm~\cite{Rangan2010} always decouples the linear mixing system~\eqref{eq:basicSystem} and~\eqref{eq:DisChannel} to parallel scalar Gaussian channels, regardless of what type of channel~\eqref{eq:DisChannel} describes the system. This feature allows more flexibility of channel types in linear mixing systems~\eqref{eq:DisChannel} than in scalar Gaussian channels~\eqref{eq:Gchannel}.

\subsection{The approximation of the minimum mean~$\ell_\infty$ error estimator}
\label{subsec:approx}
The Wiener filter is asymptotically optimal for~$\ell_\infty$ error, and one may wonder whether the performance of the Wiener filter is satisfactory for a finite signal length~$N$. Readers will see in Appendix~\ref{appen:ThmProof} that the Wiener filter is asymptotically optimal with a convergence rate on the order of~$\sqrt{\ln(N)}$, which suggests that the convergence rate is slow. Therefore, we are motivated to compare the performance of the Wiener filter with the minimum mean $\ell_p$ norm error estimators~\eqref{eq:xhat_p} in terms of the~$\ell_\infty$ norm error. Indeed, the numerical results in Section~\ref{sec:NumSim} indicate that the minimum mean~$\ell_p$ error estimator achieves a lower~$\ell_\infty$ error than the Wiener filter, provided the value of~$p$ is properly chosen. Keeping \eqref{eq:p_and_infty} in mind, one would expect that, for any positive integers $p_1$, $p_2$, where $p_1>p_2$, $\xhat_{p_1}$ always achieves a lower $\ell_\infty$ error than $\xhat_{p_2}$ does. However, experiments have indicated that, for a fixed signal dimension $N$, $\min_{\xhat_p}E[\|\xhat_{p}-\mathbf{x}\|_\infty|\mathbf{q}]$ can be achieved by a finite $p$. We include numerical results in Section~\ref{sec:NumSim}, and state our conjecture here.
\vspace*{-0.5mm}
\begin{CONJ}
Given that a system is modeled by~\eqref{eq:scalar} or~\eqref{eq:basicSystem} and~\eqref{eq:DisChannel}, where the input~${\bf x}$ is sparse Gaussian or mixture Gaussian and $\widehat{\mathbf{x}}_p$ is obtained by \eqref{eq:xhat_p}, then for any fixed signal dimension $N>0$, there exists an integer $p_\text{opt}$ such that $E[\|\widehat{\mathbf{x}}_{p_\text{opt}}-\mathbf{x}\|_\infty|\mathbf{q}]\le E[\|\widehat{\mathbf{x}}_p-\mathbf{x}\|_\infty|\mathbf{q}]$ for all positive integers $p$. Moreover, as the signal dimension $N$ increases, the value of $p_\text{opt}$ increases.
\label{con:conj01}
\end{CONJ}
\begin{REMA}
In this paper, our focus is on i.i.d. mixture Gaussian input distributions. However, additional numerical results in Section~\ref{sec:NumSim} show that Conjecture~\ref{con:conj01} also applies to other types of input distributions.
\end{REMA}

Conjecture~\ref{con:conj01} indicates that for a fixed signal dimension, the minimum mean $\ell_p$ norm error estimators~\eqref{eq:xhat_p} with different values of $p$ reduce the $\ell_\infty$ error to different amounts, and the optimal value of $p$ for a bounded signal dimension is also bounded. The conjecture also points out implicitly that $p_\text{opt}$ is a function of the signal dimension $N$. An intuitive explanation to Conjecture~\ref{con:conj01} is that as the signal dimension increases, the probability that larger errors occur also increases, and thus a larger $p$ in \eqref{eq:metricDef_lp} is used to suppress larger outliers.

\section{Numerical Results}
\label{sec:NumSim}

In this section, we provide the simulation results that inspired our results in Section \ref{subsec:approx}. Again, we first present the simulation results for parallel scalar channels, and then for linear mixing systems.

\subsection{Parallel scalar channels}
\label{subsec:SimScalar}
\begin{figure}[t]
\centering
\includegraphics[width=8.5cm]{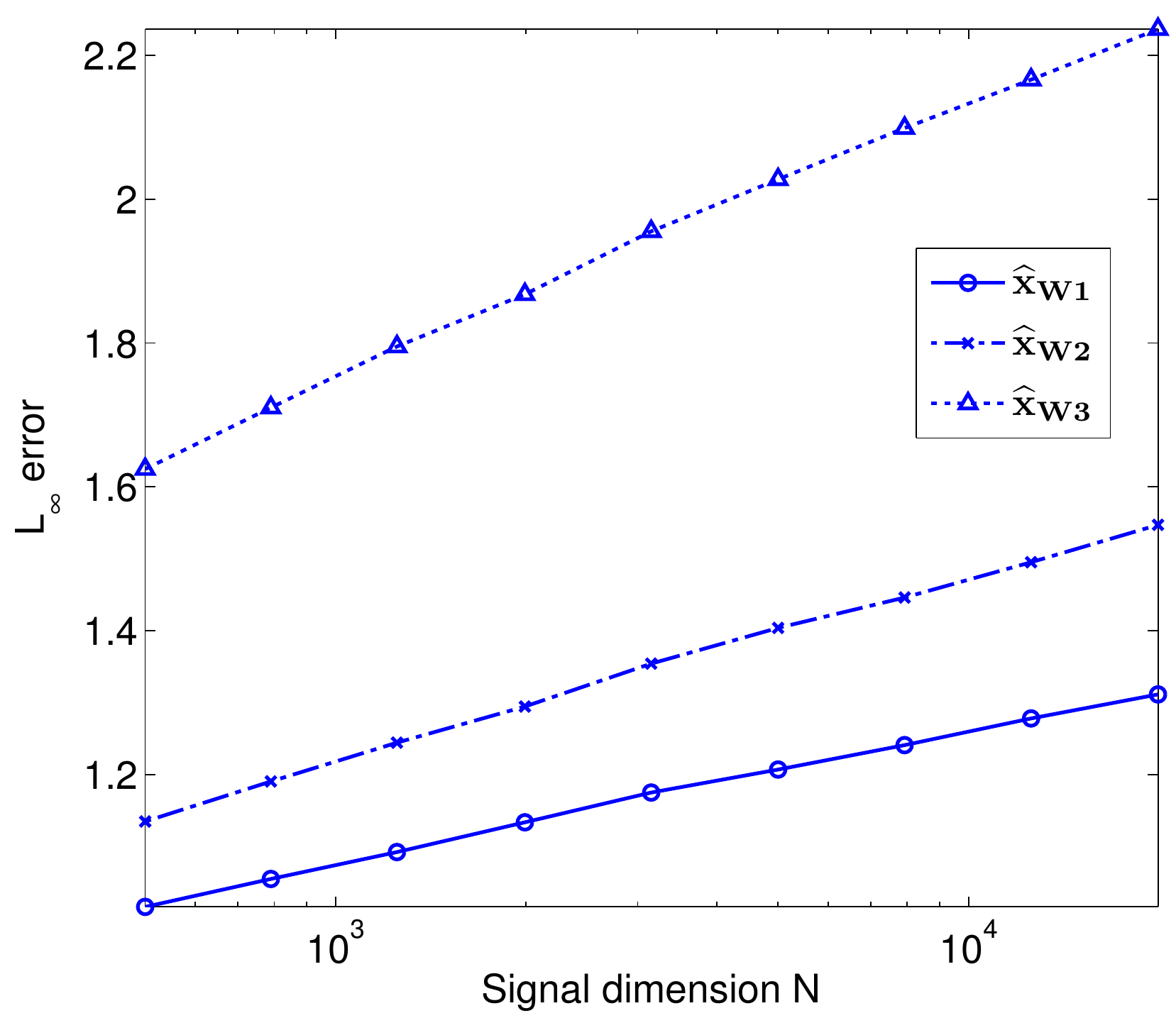}
\vspace*{-5mm}
\caption{
{\small\sl
The performance of the three Wiener filters corresponding to three different Gaussian components in parallel scalar Gaussian channels. The input vector is i.i.d.,~$x_i\sim0.2\cdot\mathcal{N}(0,10)+0.3\cdot\mathcal{N}(0,1)+0.5\cdot\mathcal{N}(0,0.5)$. The Wiener filter~${\bf\widehat{x}_{W1}}$ that corresponds to the first mixture component~$\mathcal{N}(0,10)$ achieves the lowest~$\ell_\infty$ error.
}
\label{fig:mixG}
}
\vspace*{-5mm}
\end{figure}

We first test for the parallel scalar Gaussian channels~${\bf r = x + z}$~\eqref{eq:scalar}, where the input~${\bf x}$ is i.i.d. mixture Gaussian,~$x_i\sim0.2\cdot\mathcal{N}(0,10)+0.3\cdot\mathcal{N}(0,1)+0.5\cdot\mathcal{N}(0,0.5)$, and the noise is~$z_i\sim\mathcal{N}(0,0.1)$. For this mixture Gaussian signal, there are three Wiener filters corresponding to three different input variances:~${\bf\widehat{x}}_\text{W1}=\frac{10}{10+0.1}{\bf r}$,~${\bf\widehat{x}}_\text{W2}=\frac{1}{1+0.1}{\bf r}$, and~${\bf\widehat{x}}_\text{W3}=\frac{0.5}{0.5+0.1}{\bf r}$. In Figure~\ref{fig:mixG}, we compare the~$\ell_\infty$ error of~${\bf\widehat{x}}_\text{W1}$,~${\bf\widehat{x}}_\text{W2}$, and~${\bf\widehat{x}}_\text{W3}$. It can be seen that ~${\bf\widehat{x}}_\text{W1}$, which corresponds to the Gaussian input component with largest variance, achieves the lowest~$\ell_\infty$ error among the three Wiener filters. This result verifies Theorem~\ref{coro:01}.

We then test for a set of parallel scalar Gaussian channels where the input is i.i.d. sparse Gaussian. The sparsity rate is~$s=5\%$, and the nonzero input elements are i.i.d. $\mathcal{N}(0,1)$ distributed, i.e.,~$x_i\sim\mathcal{N}(0,1)$ if~$x_i\neq0$, while the Gaussian noise is i.i.d., $z_i\sim\mathcal{N}(0,5\times 10^{-4})$; note that the signal to noise ratio (SNR) is 20dB. Here the Wiener filter is~${\bf r}/(1+5\times 10^{-4})$. We also obtain the minimum mean~$\ell_5$,~$\ell_{10}$, and~$\ell_{15}$ error estimators --~${\bf\widehat{x}}_{5}$,~${\bf\widehat{x}}_{10}$, and~${\bf\widehat{x}}_{15}$ -- using equation~\eqref{eq:xhat_p}, where we replace~${\bf q}$ by~${\bf r}$.
\begin{figure}[t]
\centering
\includegraphics[width=8.5cm]{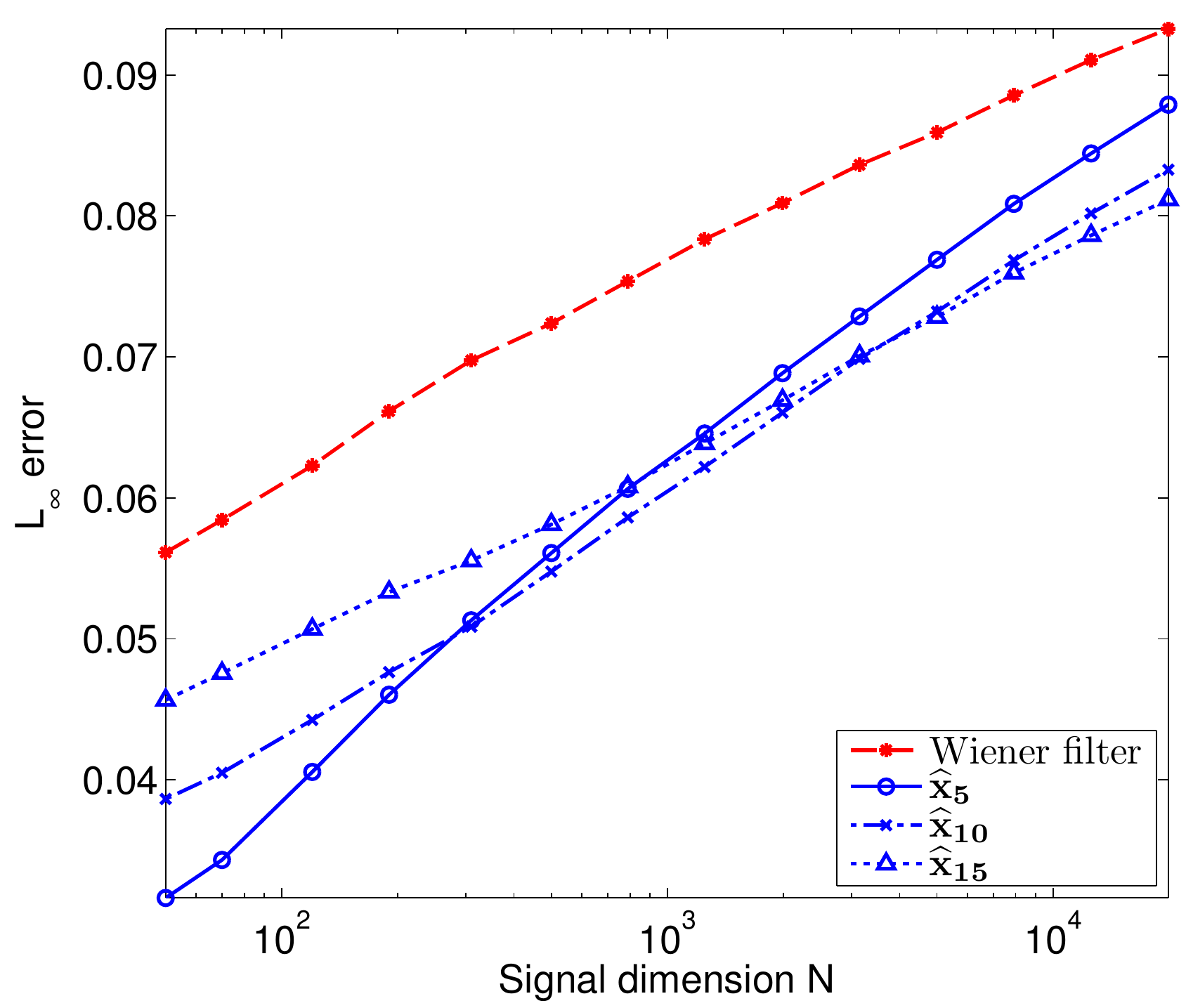}
\vspace*{-5mm}
\caption{
{\small\sl
The performance of the Wiener filter and the minimum mean $\ell_{5}$, $\ell_{10}$, and $\ell_{15}$ estimators in terms of $\ell_\infty$ error in parallel scalar Gaussian channels. The optimal $p_\text{opt}$ increases as $N$ increases. (Sparse Gaussian input, sparsity rate is~$5\%$, and SNR is 20dB.)
}
\label{fig:scalarN_VS_L_infty}
}
\vspace*{-5mm}
\end{figure}

Figure~\ref{fig:scalarN_VS_L_infty} compares the~$\ell_\infty$ error achieved by the Wiener filter and the minimum mean~$\ell_5$,~$\ell_{10}$, and~$\ell_{15}$ estimators. The results in Figure \ref{fig:scalarN_VS_L_infty} are consistent with our Conjecture \ref{con:conj01}. When $N\le300$,~$\xhat_{5}$ has the lowest~$\ell_\infty$ error among all four estimators; when~$300<N<3,000$,~$\xhat_{10}$ achieves the smallest $\ell_\infty$ error; and when~$N>3,000$,~$\xhat_{15}$ outperforms. Figure~\ref{fig:scalarN_VS_L_infty} also shows that the slope of the ``Wiener filter" line is smaller than the slopes of ``${\bf\widehat{x}}_5$", ``${\bf\widehat{x}}_{10}$", and ``${\bf\widehat{x}}_{15}$", which suggests that the Wiener filter is asymptotically optimal for~$\ell_\infty$ error.

\subsection{Linear mixing system}
We perform simulations for linear mixing systems~\eqref{eq:basicSystem} and~\eqref{eq:DisChannel} using the software package ``GAMP"~\cite{Rangan:web:GAMP} and our metric-optimal algorithm~\cite{Tan2012software}. Our metric-optimal algorithm package~\cite{Tan2012software} automatically computes equations~\eqref{eq:BayesRule}-\eqref{eq:scalarEst} where the distortion function~\eqref{eq:distDef} is given as the input of the algorithm. 

In all the following simulations, the input signals are sparse with sparsity rate~$5\%$, and the measurement matrices~$\mathbf{\Phi}$ are Bernoulli(0.5) 
and are normalized to have unit-norm rows. We have three different combinations for input distributions~\eqref{eq:basicSystem} and channel distributions~\eqref{eq:DisChannel}, whereas in all channels the SNR is 20dB:
\begin{enumerate}
\item The nonzero input entries are Gaussian~$\mathcal{N}(0,1)$, and the channel is Gaussian.

\item The nonzero input entries are Weibull distributed,
\begin{equation}
f(x_i;\lambda,k)=
\begin{cases}
\frac{k}{\lambda}\left(\frac{x_i}{\lambda}\right)^{k-1}e^{-(x_i/\lambda)^k}&x_i\ge 0\\
0&x_i<0
\end{cases},\label{eq:Weibull}
\end{equation}
where $\lambda=1$ and $k=0.5$, and the channel is Gaussian.

\item The nonzero input entries are Weibull distributed~\eqref{eq:Weibull}, and the channel is Poisson,
\begin{equation}
f_{Y|W}(y_i|w_i)=\frac{(\alpha w_i)^{y_i}e^{-(\alpha w_i)}}{y_i!},\nonumber
\end{equation}
where the scaling factor of the input is $\alpha=100$.
\end{enumerate}
For the system dimension, we fix the ratio $M/N=0.3$, and let $N$ range from $500$ to $20,000$. Then we run the Wiener filter (only in the case of sparse Gaussian input and Gaussian channel, because the Wiener filter does not apply to sparse Weibull distributed inputs), relaxed BP~\cite{Rangan2010,Rangan:web:GAMP}, and our metric-optimal algorithm with $p=5, 10, 15$ in \eqref{eq:xhat_p}. 

\begin{figure}[t]
\centering
\includegraphics[width=8.5cm]{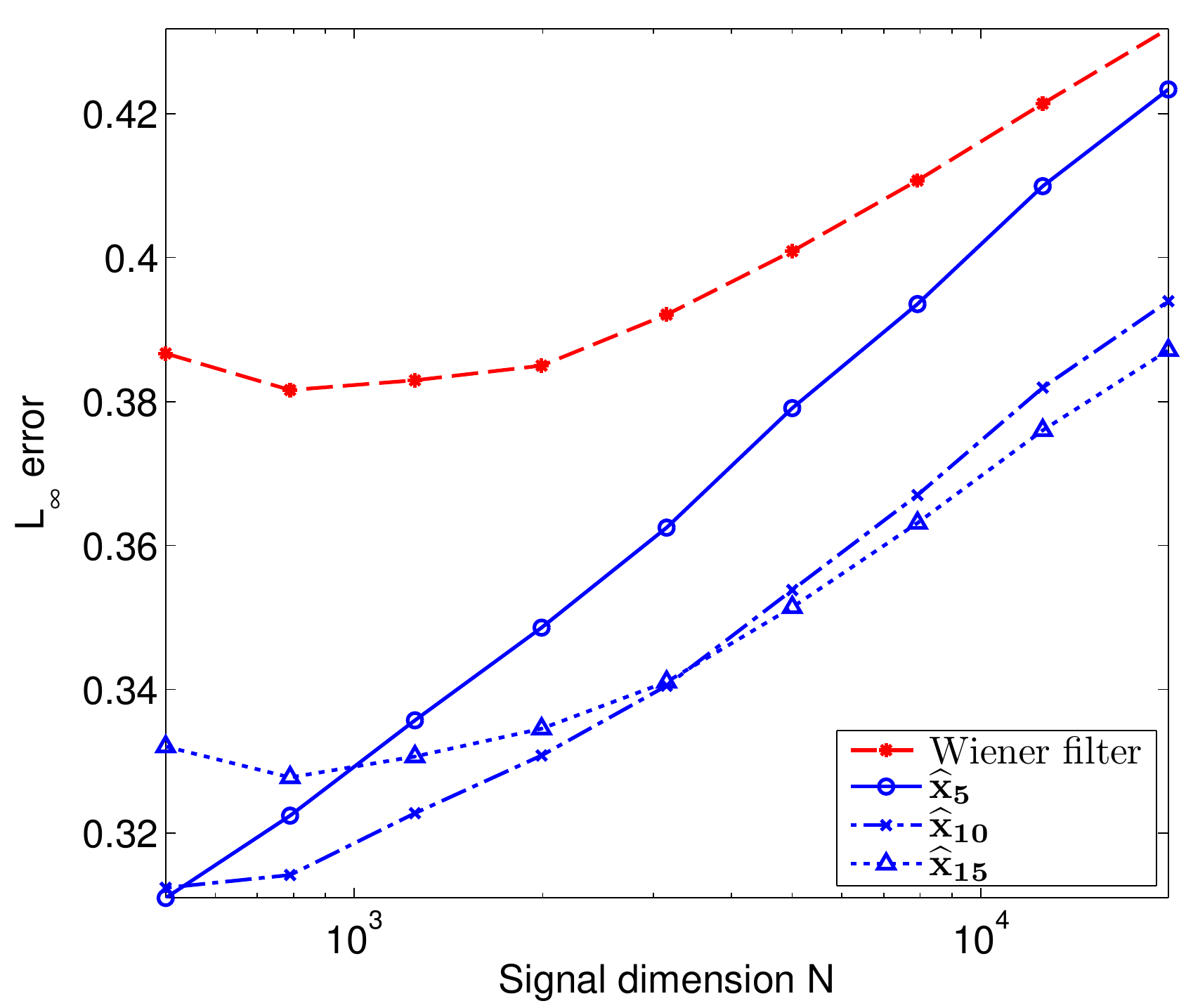} 
\vspace*{-5mm}
\caption{
{\small\sl
The performance of the Wiener filter and the minimum mean $\ell_{5}$, $\ell_{10}$, and $\ell_{15}$ estimators in terms of $\ell_\infty$ error in the linear mixing system~\eqref{eq:basicSystem} and~\eqref{eq:DisChannel}. The optimal $p_\text{opt}$ increases as $N$ increases. (Sparse Gaussian input and Gaussian channel, sparsity rate is $5\%$, and SNR=20dB.)
}
\label{fig:N_VS_L_infty}
}
\vspace*{-5mm}
\end{figure}

To compare the performance of the Wiener filter with~${\bf \widehat{x}}_p$~\eqref{eq:xhat_p}, and also to illustrate how $p_\text{opt}$ is related to the signal dimension $N$, we present in Figure~\ref{fig:N_VS_L_infty} the $\ell_\infty$ norm error of the Wiener filter, and the minimum mean $\ell_{5}$, $\ell_{10}$, and $\ell_{15}$ norm error estimators, i.e., $\xhat_{5}$, $\xhat_{10}$, and $\xhat_{15}$~\eqref{eq:xhat_p}. The numerical results shown in Figure \ref{fig:N_VS_L_infty} are similar to the results shown in Figure~\ref{fig:scalarN_VS_L_infty}, and are also consistent with Conjecture \ref{con:conj01}. 

\begin{figure}[t]
\centering
\includegraphics[width=85mm]{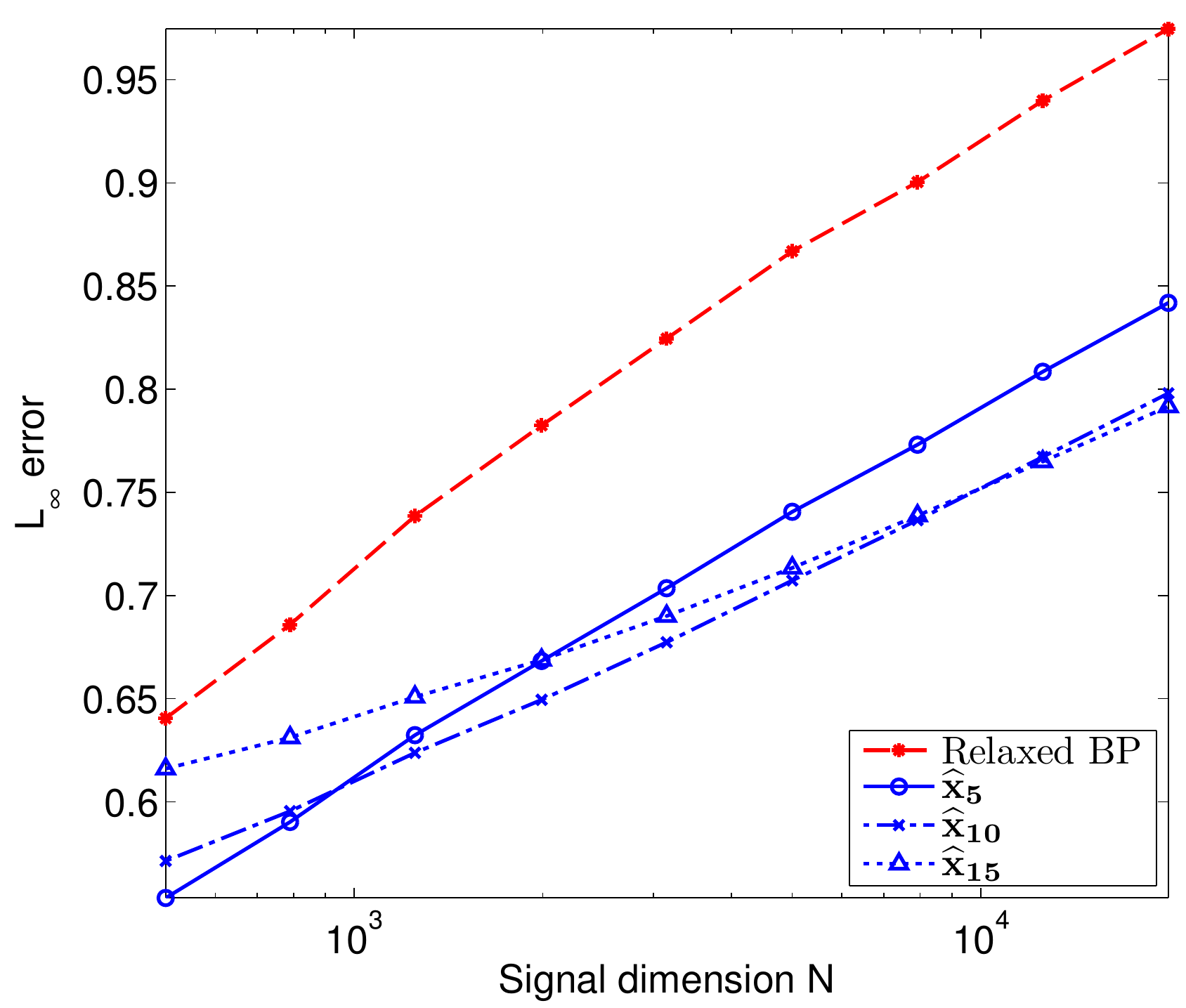}	
\caption{
{\small\sl
The performance of the relaxed BP algorithm and the minimum mean $\ell_{5}$, $\ell_{10}$, and $\ell_{15}$ estimators in terms of $\ell_\infty$ error in the linear mixing system. The optimal $p_\text{opt}$ increases as $N$ increases. (Sparse Weibull input and Gaussian channel, sparsity rate is $5\%$, and SNR=20dB.) 
}
\label{fig:Weibull_G}
}
\vspace*{-5mm}
\end{figure}

\begin{figure}[t]
\centering
\includegraphics[width=85mm]{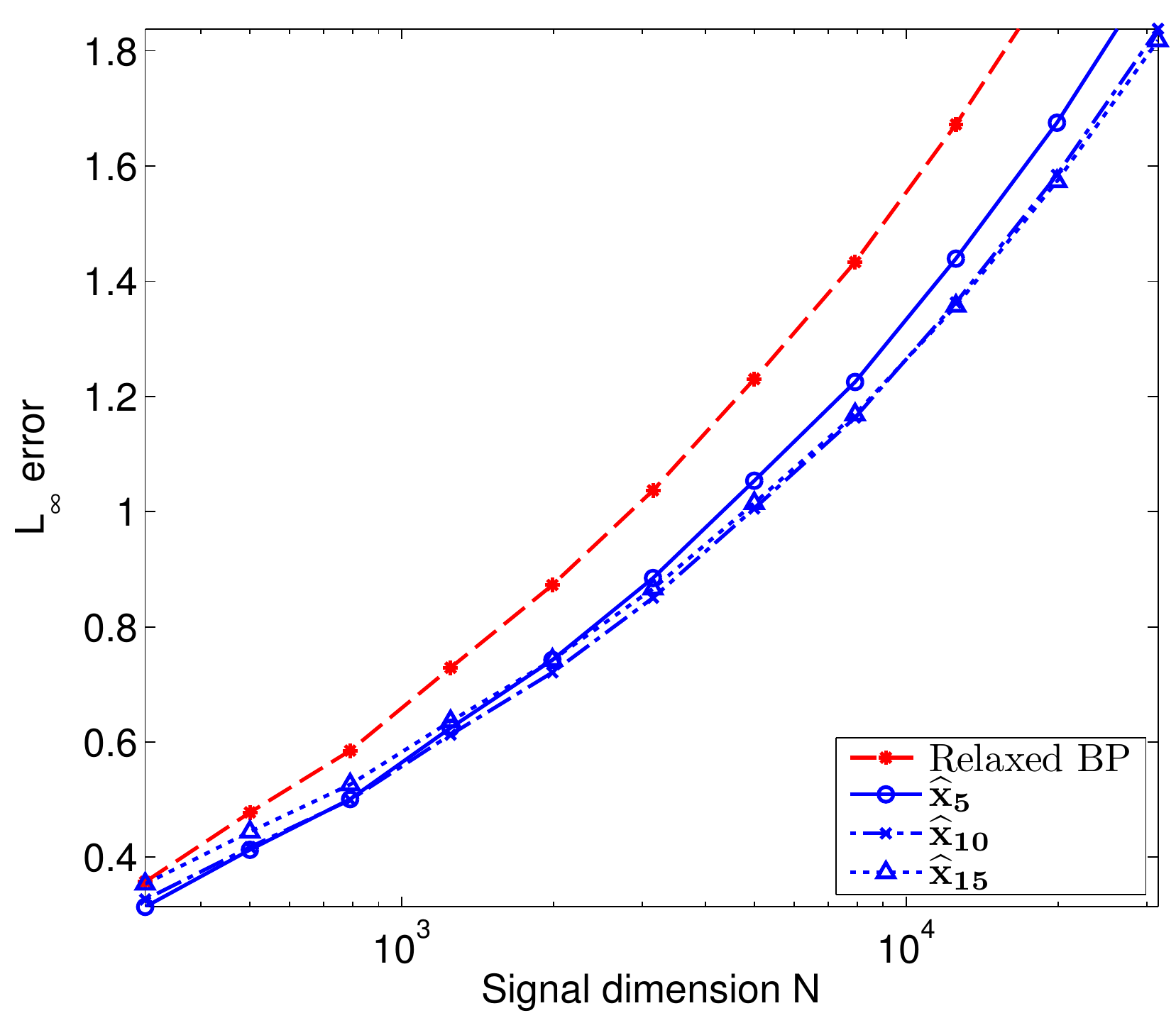}
\caption{
{\small\sl
The performance of the relaxed BP algorithm and the minimum mean $\ell_{5}$, $\ell_{10}$, and $\ell_{15}$ estimators in terms of $\ell_\infty$ error in the linear mixing system. The optimal $p_\text{opt}$ increases as $N$ increases. (Sparse Weibull input and Poisson channel, sparsity rate is $5\%$, and the SNR is 20dB.)
}
\label{fig:Weibull_P}
}
\vspace*{-5mm}
\end{figure}

When the input is sparse Weibull, the Wiener filter does not apply, because the Wiener filter is designed specifically for a Gaussian input and Gaussian channel. Instead, we compare the~$\ell_\infty$ errors of the relaxed BP, ${\bf\widehat{x}}_5$,~${\bf\widehat{x}}_{10}$, and~${\bf\widehat{x}}_{15}$~\eqref{eq:xhat_p}, and the results are shown in Figures~\ref{fig:Weibull_G} and~\ref{fig:Weibull_P}. We can see that all the minimum mean~$\ell_p$ ($p=5,10,15$) error estimators perform better than the relaxed BP algorithm for~$\ell_\infty$ error. Also, both figures suggest that the optimal $p_\text{opt}$ increases as $N$ increases, and thus the correctness of Conjecture~\ref{con:conj01} is not limited to sparse Gaussian signals.

\section{Conclusion}
\label{sec:concld}
In this paper, we studied the minimum mean~$\ell_\infty$ error estimator for both parallel scalar Gaussian channels and linear mixing systems. We showed that in both systems, when the input signal is i.i.d. sparse Gaussian or i.i.d. mixture Gaussian, the Wiener filter is asymptotically optimal for minimizing the~$\ell_\infty$ error with probability~$1$. On the other hand, when the signal dimension~$N$ is finite, our previously proposed metric-optimal algorithm with a proper~$\ell_p$ error metric outperforms the Wiener filter. A possible direction for the future work would be to find a more general form of input signal in the parallel scalar Gaussian channel setting where a linear filter is optimal for~$\ell_\infty$ norm error.

\section*{Acknowledgments}
We thank Rangan et. al. for kindly providing the Matlab code \cite{Rangan:web:GAMP} of the relaxed BP algorithm \cite{Rangan2010,RanganGAMP2010}. We also thank Nikhil Krishnan for his comments, which greatly helped us improve this manuscript.

\appendices
\section{Proof of Theorem~\ref{thm:01}}
\label{appen:ThmProof}

In order to show that the Wiener filter is optimal for $\ell_\infty$ norm error in a Gaussian channel with sparse Gaussian input, we show that the maximum absolute error caused by nonzero input elements is larger than that caused by zero elements with overwhelming probability that converges to~$1$.

Consider a set of parallel Gaussian channels~\eqref{eq:scalar}, where the input signal is $s$-sparse: $x_i\sim s\cdot\mathcal{N}(0,\mu_x)+ (1-s)\cdot\delta_0(x_i)$, and $z_i\sim\mathcal{N}(0,\mu_z)$. The Wiener filter (linear estimator) for sparse Gaussian input is $\mathbf{\widehat{x}}_\text{W,SG}=c\cdot \mathbf{r}$, and $\widehat{x}_{\text{W,SG},i}=c\cdot r_i$, where $c=\mu_x/(\mu_x+\mu_z)>0$. Let $\mathcal{I}$ denote the index set where $x_i\neq0$, i.e., $\mathcal{I}=\{i:x_i\neq0\}$, and let $\mathcal{J}$ denote the index that $\mathcal{J}=\{j:x_j=0\}$. We define two types of error patterns: ({\em i}) $x_i\neq0,i\in\mathcal{I}$, and the error $e_i=c\cdot r_i-x_i=c(x_i+z_i)-x_i=cz_i-(1-c)x_i\sim\mathcal{N}(0,c^2\mu_z+(1-c)^2\mu_x)$; ({\em ii}) ${x}_j=0,j\in\mathcal{J}$, and the error $\widetilde{e}_j=c\cdot r_j-x_j=cz_j\sim\mathcal{N}(0,c^2\mu_z)$.

It has been shown~\cite{Berman1964} that for a sequence of standard Gaussian independent random variables~$\{x_i\}_1^N$, the following equality holds:
\begin{equation}
\Pr\left(\lim_{N\rightarrow\infty}(2\ln (N))^{-1/2}\max_{1\le i\le N}x_i=1\right)=1.\label{eq:berman}
\end{equation}
Therefore, for $\mathcal{N}(0,\sigma^2)$ distributed Gaussian variable $\{x_i\}_1^N$, the equality~\eqref{eq:berman} becomes
\begin{equation}
\Pr\left(\lim_{N\rightarrow\infty}(2\ln (N))^{-1/2}\max_{1\le i\le N}\left(\frac{x_i}{\sigma}\right)=1\right)=1,\label{eq:berman_extend}
\end{equation}
or
\begin{equation}
\Pr\left(\lim_{N\rightarrow\infty}\frac{\max_{1\le i\le N}x_i}{\sigma(2\ln (N))^{1/2}}=1\right)=1.\nonumber
\end{equation}

For $e_i\sim\mathcal{N}(0,(1-c)^2\mu_x+c^2\mu_z)=\mathcal{N}(0,\sigma_1^2)$, and $\tilde{e}_j\sim\mathcal{N}(0,c^2\mu_z)=\mathcal{N}(0,\sigma_2^2)$, we get from~\eqref{eq:berman_extend} that
\begin{eqnarray}
\Pr\left\{\left.\lim_{N\rightarrow\infty}\frac{\max_{i\in\mathcal{I}} e_i}{\sigma_1\sqrt{2\ln(N(s-\epsilon))}}=1\right|{\bf r},|\mathcal{I}|=N(s-\epsilon)\right\}&=&1,\nonumber\\
\Pr\left\{\left.\lim_{N\rightarrow\infty}\frac{\max_{j\in\mathcal{J}} \tilde{e}_j}{\sigma_2\sqrt{2\ln(N(1-s+\epsilon))}}=1\right|{\bf r},|\mathcal{J}|=N(1-s+\epsilon)\right\}&=&1,\nonumber
\end{eqnarray}
where $|\mathcal{I}|$ and $|\mathcal{J}|$ denote the number of elements in the set $\mathcal{I}$ and $\mathcal{J}$, and~$\epsilon>0$ is arbitrarily small. This indicates that
\begin{equation}
\Pr\left(\left.\lim_{N\rightarrow\infty}\frac{\max_{i\in\mathcal{I}} e_i}{\max_{j\in\mathcal{J}} \widetilde{e}_j}\frac{\sigma_2\sqrt{\ln(N)+\ln(1-s+\epsilon))}}{\sigma_1\sqrt{\ln(N)+\ln(s-\epsilon))}}=1\right|{\bf r}, A_\epsilon\right)
=1,\nonumber
\end{equation}
where the event~$A_\epsilon$ is defined as
\begin{equation}
A_\epsilon=\{|\mathcal{I}|=N(s-\epsilon),|\mathcal{J}|=N(1-s+\epsilon)\},
\nonumber
\end{equation}
for some $\epsilon>0$. Note that the event $A_\epsilon$ is independent of ${\bf r}$.
Because the sparsity rate $s$ is a constant, and $\epsilon>0$ is arbitrarily small, then
\begin{equation}
\lim_{N\rightarrow\infty}\frac{\sqrt{\ln(N)+\ln(1-s+\epsilon)}}{\sqrt{\ln(N)+\ln(s-\epsilon)}}=1.\nonumber
\end{equation}
Therefore,
\begin{eqnarray}
&&\Pr\left(\left.\lim_{N\rightarrow\infty}\frac{\max_{i\in\mathcal{I}} e_i}{\max_{j\in\mathcal{J}} \widetilde{e}_j}\frac{\sigma_2\sqrt{\ln(N)+\ln(1-s+\epsilon))}}{\sigma_1\sqrt{\ln(N)+\ln(s-\epsilon))}}=1\right|{\bf r}, A_\epsilon\right)\nonumber\\
&=&\Pr\left(\left.\lim_{N\rightarrow\infty}\frac{\max_{i\in\mathcal{I}} e_i}{\max_{j\in\mathcal{J}} \widetilde{e}_j}\frac{\sigma_2}{\sigma_1}=1\right|{\bf r},A_\epsilon\right)\nonumber\\
&=&1.\nonumber
\end{eqnarray}

Because $\sigma_1>\sigma_2$, we have
\begin{equation} 
\Pr\left(\lim_{N\rightarrow\infty}\left.\max_{i\in\mathcal{I}}e_i>\lim_{N\rightarrow\infty}\max_{j\in\mathcal{J}}\widetilde{e}_j\right|{\bf r},A_\epsilon\right)=1,\nonumber
\end{equation}
i.e.,
\begin{equation} 
\Pr\left(\lim_{N\rightarrow\infty}\left.\max_{i\in\mathcal{I}}|x_i-\widehat{x}_i|>\lim_{N\rightarrow\infty}\max_{j\in\mathcal{J}}|x_j-\widehat{x}_j|\right|{\bf r},A_\epsilon\right)=1.\nonumber
\end{equation}

Recall that we denote the Wiener filter by ${\bf\widehat{x}}_\text{W,SG}$.
\begin{eqnarray}
&&E\left[\left.\lim_{N\rightarrow\infty}\|\mathbf{x}-\widehat{\mathbf{x}}_\text{W,SG}\|_\infty\right|{\bf r},A_{\epsilon}\right]\nonumber\\
&=&E\left[\left.\lim_{N\rightarrow\infty}\max_{i\in\mathcal{I}}|{ x_i-\widehat{x}_{\text{W,SG},i}}|\right|{\bf r},A_{\epsilon},\lim_{N\rightarrow\infty}\max_{i\in\mathcal{I}}|x_i-\widehat{x}_{\text{W,SG},i}|\ge\lim_{N\rightarrow\infty}\max_{j\in\mathcal{J}}|x_j-\widehat{x}_{\text{W,SG},j}|\right]\cdot\nonumber\\
&&\Pr\left(\left.\lim_{N\rightarrow\infty}\max_{i\in\mathcal{I}}|x_i-\widehat{x}_{\text{W,SG},i}|\ge\lim_{N\rightarrow\infty}\max_{j\in\mathcal{J}}|x_j-\widehat{x}_{\text{W,SG},j}\right|{\bf r},A_\epsilon\right)\nonumber\\
&&+E\left[\left.\lim_{N\rightarrow\infty}\max_{j\in\mathcal{J}}|{ x_j-\widehat{x}_{\text{W,SG},j}}|\right|{\bf r},A_{\epsilon},\lim_{N\rightarrow\infty}\max_{i\in\mathcal{I}}|x_i-\widehat{x}_{\text{W,SG},i}|<\lim_{N\rightarrow\infty}\max_{j\in\mathcal{J}}|x_j-\widehat{x}_{\text{W,SG},j}|\right]\cdot\nonumber\\
&&\Pr\left(\left.\lim_{N\rightarrow\infty}\max_{i\in\mathcal{I}}|x_i-\widehat{x}_{\text{W,SG},i}|<\lim_{N\rightarrow\infty}\max_{j\in\mathcal{J}}|x_j-\widehat{x}_{\text{W,SG},j}|\right|{\bf r},A_\epsilon\right)\nonumber\\
&=&E\left[\left.\lim_{N\rightarrow\infty}\max_{i\in\mathcal{I}}|{ x_i-\widehat{x}_{\text{W,SG},i}}|\right|{\bf r},A_{\epsilon},\lim_{N\rightarrow\infty}\max_{i\in\mathcal{I}}|x_i-\widehat{x}_{\text{W,SG},i}|\ge\lim_{N\rightarrow\infty}\max_{j\in\mathcal{J}}|x_j-\widehat{x}_{\text{W,SG},j}|\right]\cdot1\nonumber\\
&=&E\left[\left.\lim_{N\rightarrow\infty}\max_{i\in\mathcal{I}}|{ x_i-\widehat{x}_{\text{W,SG},i}}|\right|{\bf r},A_{\epsilon}\right].
\label{eq:wiener_nonzero}
\end{eqnarray}
For any estimator $\mathbf{\widehat{x}}$,
\begin{eqnarray}
&&E\left[\left.\lim_{N\rightarrow\infty}\|{\bf x-\widehat{x}}\|_\infty\right|{\bf r},A_{\epsilon}\right]\nonumber\\
&=&E\left[\left.\lim_{N\rightarrow\infty}\max_{i\in\mathcal{I\cup J}}|x_i-\widehat{x}_i|\right|{\bf r},A_{\epsilon}\right]\nonumber\\
&\ge&E\left[\left.\lim_{N\rightarrow\infty}\max_{i\in\mathcal{I}}|x_i-\widehat{x}_i|\right|{\bf r},A_{\epsilon}\right]\nonumber\\
&\ge&E\left[\left.\lim_{N\rightarrow\infty}\max_{i\in\mathcal{I}}|{ x_i-\widehat{x}_{\text{W,SG},i}}|\right|{\bf r},A_{\epsilon}\right]\label{eq:opt_VS_wiener}\\
&=&E\left[\left.\lim_{N\rightarrow\infty}\|\mathbf{x}-\widehat{\mathbf{x}}_\text{W,SG}\|_\infty\right|{\bf r},A_{\epsilon}\right].\label{eq:opt_VS_wiener02}
\end{eqnarray}
Equation~\eqref{eq:opt_VS_wiener} is true because the Wiener filter is optimal for input signals being Gaussian, and equation~\eqref{eq:opt_VS_wiener02} is true because we have shown it in~\eqref{eq:wiener_nonzero}.

We have shown that
\begin{equation}
E\left[\left.\lim_{N\rightarrow\infty}\|\mathbf{x}-\widehat{\mathbf{x}}_\text{W,SG}\|_\infty\right|{\bf r},A_{\epsilon}\right]\le E\left[\left.\lim_{N\rightarrow\infty}\|\mathbf{x}-\widehat{\mathbf{x}}\|_\infty\right|{\bf r},A_{\epsilon}\right].\nonumber
\end{equation}
It can also be shown that~\cite{Cover06} for any~$\epsilon>0$,
\begin{equation}
\Pr(A_\epsilon)\ge1-\epsilon\left|\log\left(\frac{s}{1-s}\right)\right|.\nonumber
\end{equation}
Then 
\begin{eqnarray}
&&\Pr\left(E\left[\left.\lim_{N\rightarrow\infty}\|\mathbf{x}-\widehat{\mathbf{x}}_\text{W,SG}\|_\infty\right|{\bf r}\right]\le E\left[\left.\lim_{N\rightarrow\infty}\|\mathbf{x}-\widehat{\mathbf{x}}\|_\infty\right|{\bf r}\right]\right)\nonumber\\
&=&\Pr\left(E\left[\left.\lim_{N\rightarrow\infty}\|\mathbf{x}-\widehat{\mathbf{x}}_\text{W,SG}\|_\infty\right|{\bf r},A_\epsilon\right]\le E\left[\left.\lim_{N\rightarrow\infty}\|\mathbf{x}-\widehat{\mathbf{x}}\|_\infty\right|{\bf r},A_\epsilon\right]\right)\Pr\left(A_\epsilon\right)\nonumber\\
&&+\Pr\left(E\left[\left.\lim_{N\rightarrow\infty}\|\mathbf{x}-\widehat{\mathbf{x}}_\text{W,SG}\|_\infty\right|{\bf r},A_\epsilon^c\right]\le E\left[\left.\lim_{N\rightarrow\infty}\|\mathbf{x}-\widehat{\mathbf{x}}\|_\infty\right|{\bf r},A_\epsilon^c\right]\right)\Pr\left(A_\epsilon^c\right)\nonumber\\
&=&1\cdot\Pr\left(A_\epsilon\right)+\Pr\left(E\left[\left.\lim_{N\rightarrow\infty}\|\mathbf{x}-\widehat{\mathbf{x}}_\text{W,SG}\|_\infty\right|{\bf r},A_\epsilon^c\right]\le E\left[\left.\lim_{N\rightarrow\infty}\|\mathbf{x}-\widehat{\mathbf{x}}\|_\infty\right|{\bf r},A_\epsilon^c\right]\right)\Pr\left(A_\epsilon^c\right)\nonumber\\
&\ge&1\cdot\Pr\left(A_\epsilon\right)\nonumber\\
&\ge&1-\epsilon\left|\log\left(\frac{s}{1-s}\right)\right|.\nonumber
\end{eqnarray}
Therefore, $\widehat{\mathbf{x}}_\text{W,SG}$ is asymptotically optimal for $\ell_\infty$ norm error with probability $1$ when~$\epsilon\rightarrow0$.

\section{Proof of Theorem~\ref{coro:01}}
\label{appen:coro_01}
The input signal of the scalar Gaussian channels~\eqref{eq:scalar} is i.i.d. mixture Gaussian, $x_i\sim s_1\cdot\mathcal{N}(0,\mu_1)+s_2\cdot\mathcal{N}(0,\mu_2)+\cdots+s_K\cdot\mathcal{N}(0,\mu_K)$, and suppose without loss of generality that~$\mu_1=\max_{k\in\{1,2,\ldots,K\}}\mu_k$. The Wiener filter is $\mathbf{\widehat{x}}_\text{W,MG}=c\cdot \mathbf{r}$, and $\widehat{x}_{\text{W,MG},i}=c\cdot r_i$, where $c=\mu_1/(\mu_1+\mu_z)>0$ is a constant. Let $\mathcal{I}_k$ denote the index set where $x_i\sim\mathcal{N}(0,\mu_k)$, i.e.,~$\mathcal{I}_k=\{i:x_i\sim\mathcal{N}(0,\mu_k)\}$ for~$k\in\{1,2,\ldots,K\}$. Then we define~$K$ types of error patterns: $x_i\sim\mathcal{N}(0,\mu_k)$, and the error $e_{k,i}=c\cdot r_i-x_i=c(x_i+z_i)-x_i=cz_i-(1-c)x_i\sim\mathcal{N}(0,c^2\mu_z+(1-c)^2\mu_k)$. Because the noise variance~$\mu_z$ is a constant, we have
\begin{equation}
\max_{k\in\{1,2,\ldots,K\}}(c^2\mu_z+(1-c)^2\mu_k)=(c^2\mu_z+(1-c)^2\mu_1).\nonumber
\end{equation}

Define the event~$A_\epsilon$ as
\begin{equation}
A_\epsilon=\{|\mathcal{N}_1|=N(s_1+\epsilon_1),|\mathcal{N}_2|=N(s_2+\epsilon_2),\ldots,|\mathcal{N}_K|=N(s_K+\epsilon_K)\},\nonumber
\end{equation}
where~$\sum_{k=1}^K\epsilon_k=0$.

Again applying equation~\eqref{eq:berman} and following the same procedures in the proof of Theorem~\ref{thm:01}, we get that
\begin{equation}
\Pr\left(\left.\max_{i\in\mathcal{I}_1}e_{1,i}\ge\max_{i\in\mathcal{I}_k}e_{k,i}\right|{\bf r},A_\epsilon\right)=1,\nonumber
\end{equation}
for any~$k\in\{2,3,\ldots,K\}$. Then applying the same derivation as equations~\eqref{eq:wiener_nonzero} and~\eqref{eq:opt_VS_wiener02}, we have
\begin{eqnarray}
E\left[\left.\lim_{N\rightarrow\infty}\|\mathbf{x}-\widehat{\mathbf{x}}_\text{W,MG}\|_\infty\right|{\bf r},A_{\epsilon}\right]
=E\left[\left.\lim_{N\rightarrow\infty}\max_{i\in\mathcal{I}_1}|{ x_i-\widehat{x}_{\text{W,MG},i}}|\right|{\bf r},A_{\epsilon}\right],\nonumber
\end{eqnarray}
and
\begin{eqnarray}
E\left[\left.\lim_{N\rightarrow\infty}\|{\bf x-\widehat{x}}\|_\infty\right|{\bf r},A_{\epsilon}\right]
\ge E\left[\left.\lim_{N\rightarrow\infty}\|\mathbf{x}-\widehat{\mathbf{x}}_\text{W,MG}\|_\infty\right|{\bf r},A_{\epsilon}\right]\nonumber
\end{eqnarray}
for any estimand~${\bf \widehat{x}}$.
Because for any~$\epsilon_1,\epsilon_2,\ldots,\epsilon_K>0$~\cite{Cover06},
\begin{equation}
\Pr(A_\epsilon)\ge1-\sum_{k=1}^K\epsilon_k\left|\log(s_k)\right|,\nonumber
\end{equation}
 and finally we get
\begin{equation}
\Pr\left(E\left[\left.\lim_{N\rightarrow\infty}\|\mathbf{x}-\widehat{\mathbf{x}}_\text{W,MG}\|_\infty\right|{\bf r}\right]\le E\left[\left.\lim_{N\rightarrow\infty}\|\mathbf{x}-\widehat{\mathbf{x}}\|_\infty\right|{\bf r}\right]\right)
\ge1-\sum_{k=1}^K\epsilon_k\left|\log(s_k)\right|.\nonumber
\end{equation}

\ifCLASSOPTIONcaptionsoff
  \newpage
\fi



\bibliography{cites}

\end{document}